\documentclass{ws-ijmpa}
\usepackage{amsmath,amsfonts,amssymb}
\begin{document}

\markboth{S. S. Afonin, A. A. Andrianov}{Matching to OPE}

\catchline{}{}{}{}{}

\title{Matching meson resonances to OPE in QCD}
\author{A.~A. Andrianov}
\address{INFN, Sezione di Bologna, Bologna 40126, Italy\\
St.~Petersburg State University, St.~Petersburg 198905, Russia} 
\author{S.~S. Afonin and D. Espriu}
\address{Departament d'ECM, 
Universitat de Barcelona, \\
Barcelona 08028, Spain}
\author{V.~A. Andrianov}
\address{St.~Petersburg State University, St.~Petersburg 198905, Russia}

\maketitle


\begin{abstract}
We investigate the possible corrections to the linear Regge trajectories
 for the light-quark meson sector by
matching two-point correlators of quark currents to the Operator Product
Expansion. We find that the allowed modifications to the linear
behavior must decrease rapidly with the principal quantum number. After
fitting the lightest states in each channel and certain low-energy constants
the whole spectrum for meson masses and residues is obtained in a satisfactory
agreement with phenomenology. The perturbative corrections to our results are discussed.
\end{abstract}


The observed masses squared of mesons with given quantum numbers
form linear trajectories~\cite{bgs,ani} depending on the number of
radial excitation $n$. This is a strong indication that QCD admits
an effective string description, as this type of spectrum is
characteristic e.g. of the bosonic string. In the bosonic string model
the slope of all
trajectories must be equal since this quantity is proportional to
the string tension depending on gluedynamics only. However,
there exist sizeable deviations
from the string picture. In the
present analysis we examine possible corrections to the linear
trajectories in the vector~(V), axial-vector~(A), scalar~(S), and
pseudoscalar~(P) channels~\cite{jhep}. Our method is based on the
consideration of the two-point correlators of V,A,S,P quark
currents in the large-$N_c$ limit of QCD~\cite{hoof}. On the one hand,
by virtue of confinement they are saturated by an infinite set of
narrow meson resonances, that is,
they can be represented by the sum of related meson poles in
Euclidean space:
\begin{gather}
\label{cor1}
\Pi_{J}(Q^{2})=\int d^{4}x\exp(iQx)\langle\bar{q}\Gamma
q(x)\bar{q}\Gamma q(0)
\rangle_{\mbox{\scriptsize{planar}}}=\sum_n\frac{2F_J^2(n)}{Q^2+m_J^2(n)},
\end{gather}
expressing the quark-hadron duality\cite{sh}.
Here $J\equiv S,P,V,A$; $\Gamma=i,\gamma_{5},\gamma_{\mu},\gamma_{\mu}\gamma_{5}$. Further we denote $F_{S,P}\equiv G_{S,P}m_{S,P}$.
On the other hand, their high-energy asymptotics is
provided by the perturbation theory and the Operator Product Expansion (OPE) with
 condensates~\cite{svz}. 
Matching these two approaches results in the
Chiral Symmetry Restoration sum rules representing
a set of constraints on meson mass parameters~\cite{aaa}. 
We performed the further analysis in the chiral limit and firstly 
in the leading order of perturbation theory.

Let us consider the linear ansatz for the meson mass spectra with a
non-linear correction $\delta$:
\begin{equation}
\label{acor1}
m^2_{J}(n)=m_{0,J}^2+a\,n+\delta_{J}(n), \qquad J\equiv V,A,S,P.
\end{equation}
The last term in Eq.~(\ref{acor1}) signifies a possible
deviation from the string picture in QCD. The condition of
convergence for the generalized Weinberg sum rules~\cite{beane} imposes the
equality of slopes and a falloff of non-linear
correction.

The asymptotic freedom leads to the relation between residues and
masses: $F^2(n)\sim \frac{dm^2(n)}{dn}$. Our analysis showed that
the analytical structure of the OPE admits, however, exponentially
small deviations from this relation (in contrast to to~\cite{beane,simon}),
\begin{gather}
m^2_{J}(n)=M_{J}^2+an+A_m^{J}e^{-B_m\cdot n}, \\
F^2_{V\!,A}(n)=a\left(\frac{1}{8\pi^2}+A_F^{V\!,A}e^{-B_F\cdot n}\right),\qquad
G^2_{S,P}(n)=a\left(\frac{3}{16\pi^2}+A_G^{S,P}e^{-B_G\cdot n}\right)
\end{gather}
with certain constants $A_{m,F,G}^{V,A,S,P}$ and $B_{m,F,G}>0$ to be fitted.
It is plausible to suppose that, for masses, the dynamics under these
exponentially small corrections 
is governed
mostly by gluons and thereby does not
depend on flavor. Thus, we keep the exponent $B_m$ the same for
all channels. For the same reason we regard $B_{F,G}$ as independent
of parity.

In Tables 1 and 2 we show an example of numerical fits resulting from our approach where in the S,P-channels two possibilities are considered: $\pi$-meson belongs to the radial Regge trajectory ('$\pi$-in') and does not ('$\pi$-out'). The inputs general for all tables (if any)
are: $a=(1120\,\mbox{MeV})^2$, $\langle\bar
qq\rangle=-(240\,\mbox{MeV})^3$,
$\frac{\alpha_s}{\pi}\langle\left(G_{\mu\nu}^a\right)^2\rangle=
(360\,\mbox{MeV})^4$, $f_{\pi}=103\,\mbox{MeV}$,
$Z_{\pi}=2\frac{\langle\bar qq\rangle^2}{f_{\pi}^2}$,
$\alpha_s=0.3$. The units are: $m(n)$, $F(n)$, $G(n)$ --- MeV; $A_{m}$
--- MeV${}^2$; $A_F$, $A_G$, $B_{F,G,m}$ --- MeV${}^{0}$.

Let us discuss the impact of running coupling constant at next-to-leading order of perturbation theory.
We have in this case the contribution to the imaginary part of correlator which is related to the full correlator through the dispersion relation. In the large-$N_c$ limit for the vector and axial-vector case this leads to:
\begin{equation}
\label{pt1}
{\text Im}\Pi(t)=\frac{1}{4\pi^2}\left(1+\frac{\alpha_s(t)}{\pi}\right)+\mathcal{O}(\alpha_s^2)
=\pi\sum_{n=0}^{\infty}2F^2(n)\delta\left(t-m^2(n)\right).
\end{equation}
We perform summation in Eq. \eqref{pt1} by applying the Euler-Maclaurin summation formula. This provides smoothness of this expression. Finally we obtain the relation:
\begin{equation}
\label{pt4}
F^2(n_0)=\frac{dm^2(n)}{dn}\frac{1}{8\pi^2}\left(1+\frac{\alpha_s\left(t(n_0)\right)}{\pi}\right).
\end{equation}
\begin{table}[t]
\tbl{An example of parameters for the mass spectra. 
The existing experimental values~$^{2,11}$
are displayed in brackets.}
{\footnotesize
\begin{tabular}{@{}ccc@{}} \toprule
  Case &  Inputs & Fits and constants \\\colrule
   $VA$ & {\begin{tabular}{l}
   $m_V(0)=770\,(769.3\pm0.8)$,\\
   $m_A(0)=1200\,(1230\pm40)$,\\
   \end{tabular}}
   & {\begin{tabular}{l}
   $M=920$, $B_m=0.97$, $B_F=0.72$,\\
   $A_m^V=-500^2$, $A_m^A=770^2$,\\ $A_F^V=0.0012$,
   $A_F^A=-0.0031$,\\ $L_{10}=-6.5\cdot10^{-3}$, $\Delta m_{\pi}=2.3$
   \end{tabular}} \\\colrule
  {\begin{tabular}{c}
   $SP$\\
   ($\pi$-in)
   \end{tabular}}
   & {\begin{tabular}{l}
   $m_S(0)=1000$,\\
   $m_P(0)=0$,\\
   $m_P(1)=1300\,(1300\pm100)$,\\
   $B_m=0.97$
   \end{tabular}}
   & {\begin{tabular}{l}
   $\bar{M}=840$, $B_G=0.42$,\\ $A_m^S=550^2$,
   $A_m^P=-840^2$, \\$A_G^S=-0.0009$, $A_G^P=0.0004$, \\
   $L_8=1.0\cdot10^{-3}$
   \end{tabular}} \\\colrule
   {\begin{tabular}{c}
   $SP$\\
   ($\pi$-out)
   \end{tabular}}
   & {\begin{tabular}{l}
   $m_S(0)=1000$,\\
   $m_P(0)=1300\,(1300\pm100)$,\\
   $m_P(1)=1800\,(1801\pm13)$,\\
   $B_m=0.97$
   \end{tabular}}
   & {\begin{tabular}{l}
   $\bar{M}=1470$, $B_G=1.27$,\\ $A_m^S=-1080^2$,
   $A_m^P=-690^2$,\\ $A_G^S=0.0213$, $A_G^P=0.0067$,\\
   $L_8=0.9\cdot10^{-3}$
   \end{tabular}} \\
  \botrule
\end{tabular}}
\end{table}
\begin{table}[t]
\tbl{Mass spectrum and residues for the parameter sets of
Table~1.}
{\footnotesize
\begin{tabular}{@{}cccccc@{}}\toprule
  $n$ & out & $0$ & $1$ & $2$ & $3$ \\\colrule
 $m_V(n)$  & & $770\,(775.8\pm0.5)$ & $1420\,(1465\pm25)$
& $1820\,(1900?)$ &
  $2140\,(2149\pm17)$ \\
  $F_V(n)$ &  & $138\,(154\!\pm\!8)$ & $135$ &  $133$  & $133$ \\
  $m_A(n)$ & & $1200\,(1230\pm40)$& $1520\,(1640\pm40)$ &
  $1850\,(1971\pm15)$ & $2150\,(2270\pm50)$\\
  $F_A(n)$ &  & $116\, (123\pm25)$ & $125$ & $128$ &  $130$  \\
  \colrule
  $m_S(n)$ &  & $1000 \,(980\pm10)$ & $1440\,(1507\pm5)$ & $1800\, (1714\pm5)$
& $2100$ \\
  $G_S(n)$ &  & $176$ & $178$ & $178$ & $179$ \\
  $m_P(n)$ &  & $0$ ($\pi$-in) & $1300\,(1300\pm100)$ &$1760\,(1801\pm13)$ &
  $2100\,(2070\pm35)$ \\
  $G_P(n)$ &  & -- & $179$ & $179$ & $179$  \\
  \colrule
  $m_S(n)$ &  & $1000\, (980\pm10)$ & $1730\, (1714\pm5)$
& $2120$ & $2420$ \\
  $G_S(n)$ &  & $243$ & $199$ & $185$ & $181$ \\
  $m_P(n)$ & $0$ & $1300\,(1300\pm100)$ & $1800\,(1812\pm14)$ &
  $2150\,(2070\pm35)$ & $2430\,(2360\pm30)$ \\
  $G_P(n)$ & -- & $201$ & $186$ & $181$ & $180$  \\\botrule
\end{tabular}}
\end{table} 
Eq. \eqref{pt4} can be approximated by finite differences. For the first two states this reads:
\begin{equation}
\label{pt5}
\frac{F_{\rho}^2}{m_{\rho'}^2-m_{\rho}^2}\approx\frac{F_{a_1}^2}{m_{a_1'}^2-m_{a_1}^2}\approx
\frac{1}{8\pi^2}\left(1+\frac{\alpha_s(1\,\text{GeV})}{\pi}\right).
\end{equation}
Substituting experimental values \cite{pdg} into Eq. \eqref{pt5} one arrives at the estimate: 
$1.5\pm0.2\approx1.3\pm0.6\approx1.4$
in a good agreement with the phenomenology.

In order to reproduce the running coupling behaviour in Eq. \eqref{pt1} one should accept the following ansatz for the residues:
\begin{equation}
\label{pt6}
F^2(n)\simeq \frac{dm^2(n)}{dn}\frac{1}{8\pi^2}\left(1+4\left(\beta_0\ln\frac{m^2(n)}{\Lambda^2_{\text{QCD}}}\right)^{-1}\right)
+\text{exp. corr.}
\end{equation}
However this ansatz introduces into the OPE the additional terms $\sim \frac{\ln Q^2}{Q^2}$ (and powers of these terms) which are absent in the standart OPE. Thus ansatz \eqref{pt5} must be improved. This problem is under our studies now. 

Let us summarize the results of our analysis:\\
{\bf 1)} The convergence of the generalized Weinberg sum rules requires
the universality of slopes and intercepts for parity conjugated trajectories.\\
{\bf 2)} The matching to the OPE cannot be achieved by a simple linear
parameterization of the mass spectrum, the linear trajectory ansatz. 
There must exist deviations from the
linear trajectory ansatz triggered by  chiral
symmetry breaking. These deviations must decrease at least
exponentially with $n$.\\
{\bf 3)} For heavy states, the D-wave vector mesons have to decouple from
asymptotic sum rules. This fact implies the exponential (or faster)
decreasing the corresponding decay constants $F^2_D(n)$.\\
{\bf 4)} Our results seem to exclude a light $\sigma(600)$ particle as a quarkonium state and
rather favor the non-linear realization of chiral symmetry with the lightest
scalar of  mass $\sim 1$ GeV, its chiral partner being the $\pi'(1300)$.\\
{\bf 5)} In our approach the quantities $L_8, L_{10}$ and $\Delta m_\pi$
are obtained, in  satisfactory agreement with the phenomenology.\\
{\bf 6)} Perturbative corrections can be systematically treated making, nevertheless, small effects on the fits presented.

\vspace{-2mm}
\section*{Acknowledgements}

This work is supported by RFBR, grant 05-02-17477, grant UR 02.01.299 and 
INFN, grant IS/PI13. Work of S. A. and D. E. is partially supported by CYT FPA, grant 2004-04582-C02-01 and by CIRIT GC, grant 2001SGR-00065.

\end{document}